\documentclass[useAMS,usenatbib,onecolumn,usegraphicx]{mn2e}
\usepackage{amssymb}

\title[Hall equilibrium of thin Keplerian disks ]{Hall equilibrium of thin Keplerian disks embedded in mixed poloidal and toroidal magnetic fields}
\author[Yuri M. Shtemler, Michael Mond, and G$\ddot{u}$nther R$\ddot{u}$diger]{Yuri M. Shtemler$^{1}$\thanks{E-mail:
shtemler@bgu.ac.il; mond@bgu.ac.il; gruediger@aip.de},  Michael Mond$^{1}$, and G$\ddot{u}$nther R$\ddot{u}$diger$^{2}$\\
$^{1}$Department of Mechanical Engineering,  Ben-Gurion
University of the Negev,  P.O. Box 653, Beer-Sheva 84105,
Israel\\
$^{2}$Astrophysikalisches Institut Potsdam, An der Sternwarte 16,
D-14482 Potsdam, Germany
}
\begin{document}

\date{Accepted ---. Received ----; in original form ----}

\pagerange{\pageref{firstpage}--\pageref{lastpage}} \pubyear{}

\maketitle

\label{firstpage}

\begin{abstract}
Axisymmetric steady-state  weakly ionized Hall-MHD Keplerian thin
disks are investigated by using asymptotic expansions in the small
disk aspect ratio $\epsilon$. The model incorporates the azimuthal
and poloidal components of the magnetic fields in the leading
order in $\epsilon$. The disk structure is described by an
appropriate Grad-Shafranov equation for the poloidal flux function
$\psi$ that involves two arbitrary functions of $\psi$ for the
toroidal and poloidal currents. The flux function is symmetric
about the midplane and satisfies certain boundary conditions at
the near-horizontal disk edges. The boundary conditions model the
combined effect of the primordial as well as the  dipole-like
magnetic fields. An analytical solution for the Hall equilibrium
is achieved by further expanding the relevant equations in an
additional small parameter $\delta$ that is inversely proportional
to the Hall parameter. It is thus found that the Hall equilibrium
disks fall into two types: Keplerian disks with (i) small
($R_d\sim\delta^0$) and (ii) large (${R_d}\gtrsim\delta^{-k}$,
$k>0$)  radius of the disk. The numerical examples that are
presented demonstrate the richness and great variety of magnetic
and density configurations that may be achieved under the Hall-MHD
equilibrium.
\end{abstract}


\section{Introduction}

In the present study the Hall equilibrium of thin Keplerian disks embedded in 3D axially-symmetric magnetic fields is investigated. Protoplanetary disks have been the subject of numerous investigations in search of possible instabilities that may play a significant role in such yet not totally understood phenomena as the outwards transfer of angular momentum, and planet formation. However, in most astrophysics-related hydromagnetic stability studies which include accretion, jet, and wind effects the models for the underlying equilibrium states are based on the local approximation or a 'cylindrical disk' that is uniform in the axial direction.  A more realistic model of thin disks is adopted in \cite{Regev 1983}, \cite{Kluzniak and Kita 2000}, \cite{Umurhan et al. 2006} for equilibrium nonmagnetized rotating disks (see also \cite{Ogilvie 1997} for Keplerian disks of ideal MHD plasmas) and is based on an asymptotic approach in the small aspect ratio of the disk.  Traditionally, a big variety of geometrical configurations of axially symmetric steady-state equilibria are described through a few arbitrary functions of the poloidal magnetic flux. The latter satisfies the Grad-Shafranov (GS) type equations within both one-fluid and multifluid models of the plasmas (e.g. \cite{McClements and Thyagaraja 2001, Thyagaraja and McClements 2006}, see also the earlier study by \cite{Lovelace et al. 1986}, where in particular thin rotating disks have been discussed and references therein). Such solutions provide a base for their stability studies.

Although the magnetic field configuration in real disks is largely unknown and, as observations and numerical simulations indicate, toroidal magnetic component may be of the same order of magnitude, or some times dominate the poloidal field due to the differential rotation of the disk (see e.g. \cite{Terquem and Papaloizou 1996, Papaloizou and Terquem 1997, Hawley and Krolik 2002, Proga 2003}).
As shown in the present study there are three kinds of  possible  equilibrium in thin rotating disks which can not be reduced from one to the other: (i) pure poloidal, (ii) pure toroidal and (iii) mixed toroidal and poloidal magnetic field equilibria. The first case has been considered within the ideal MHD model by \cite{Lovelace et al. 1986}, and \cite{Ogilvie 1997}. The second case, i.e. the pure toroidal magnetic field has been investigated within the Hall MHD model by \cite{Shtemler et al. 2007}. The present study is aimed to describe numerous axially-symmetric equilibria in 3D axially-symmetric magnetic fields that contain both toroidal and poloidal components, i.e. case (iii) within the Hall MHD model in thin disk approximation.

While most of recent research has focused on the
magnetohydrodynamic (MHD) description of magneto-rotational
instability (MRI), the importance of the Hall electric field to
such astrophysical objects as protoplanetary disks has been
pointed out by Wardle (1999). Since then, such works as
\cite{Balbus and Terquem 2001}, \cite{Salmeron and Wardle 2003},
\cite{Salmeron and Wardle 2005}, \cite{Desch 2004}, \cite{Urpin
and Rudiger 2005}, \cite{Rudiger and Kitchatinov 2005}, and
\cite{Pandey and Wardle 2008} have shed light on the role of the
Hall effect in the modification of the MRIs. However, in addition
to modifying the MRIs, the Hall electromotive force gives rise to
a new family of instabilities in a density-stratified environment.
That new family of instabilities is characterized by the Hall
parameter  $\pi_H$ of the order of unity ( $\pi_H$  is the ratio
of the Hall drift velocity  $V_{HD}$ to the characteristic
velocity, where $V_{HD}=l_i^2\Omega_i/H_*$,  $l_i$ and  $\Omega_i$
are the inertial length and Larmor frequency of the ions,
respectively, and  $H_*$  is the density inhomogeneity length (see
\cite{Huba 1991}). As shown by \cite{Liverts and Mond 2004}, and
\cite{Kolberg et al. 2005}, the Hall electric field combined with
radial density stratification gives rise to two modes of wave
propagation. The first one is the stable fast magnetic penetration
mode, while the second is a slow quasi-electrostatic mode that may
become unstable for a short-enough density inhomogeneity length
$H_*$. The latter is termed the Hall instability. Density
stratifications also play a crucial role in the recently discussed
starto-rotational instability. That instability is excited by the
combined effect of vertical and radial density stratification.

The aim of the present work is, therefore to describe the various Hall equilibrium configurations of magnetic field and density in thin magnetized Keplerian disks as a first step towards a further study of their various density stratification-dependent instabilities.
The following simplifying assumptions are adopted throughout the present work: the plasma is treated in a 3D axially symmetric magnetic field within the Hall MHD model.  The problem inside the disk is treated within the thin-disk approximation, while the effects of the outer region is modeled by the boundary condition for the poloidal magnetic flux. Influence of the accretion, jet and wind effects are neglected in the thin-disk approximation (\cite{Ogilvie 1997}). A near Keplerian region of the rotating disk is considered that starts at some radial distance away from a central body.  The influence of the latter on the Keplerian portion of the disk is modeled by a dipole-like contribution (singular at the disk axis) to the boundary condition for the poloidal magnetic flux at the near-horizontal disk edges. In addition, axial electrical currents that are localized in the central non-Keplerian regions of the disk are taken into account through the resulting toroidal magnetic field components within the rotating disk.  Influence of the interstellar (primordial) magnetic field is modeled by specifying the contribution of the latter on the near-horizontal disk edges.

The paper is organized as follows. The dimensional governing equations, the normalization procedure, and the resulting non-dimensional system are presented in the next Section. Section 3 contains an asymptotic analysis of the Hall-equilibrium in the small aspect ratio as well as the derivation the appropriate GS equation for the flux function and the corresponding boundary conditions at the disk edge. In section 4 an analytic model is developed for the Hall equilibrium with the aid of asymptotic expansions in a newly defined small parameter inversely proportional to the Hall parameter. Numerical examples for different regimes of the equilibrium are presented in section 5. Summary and discussion are in section 6.

\section{The physical model for THIN Keplerian DISKs}
A rotating thin disks is considered that is under the influence of a general 3D axially-symmetric magnetic field, as well as the gravitational field due to a massive central body. Electron inertia and pressure, displacement current, viscosity and radiation effects are neglected in the electrons momentum equation. Consequently the latter is reduced to the generalized Ohm law that takes into account the Hall effect. In addition the plasma is assumed to be quasi-neutral.

\subsection{The basic equations}
Under the assumptions mentioned above the physical model of the rotating magnetized disk is given by:
\begin{equation}
\frac{\partial n} {\partial t}
               + \nabla \cdot(n {\bf{V}} ) =0,\\
\label{1}
\end{equation}
\begin{equation}
m_{i}n \frac{ D \bf {V} }{Dt} = -\nabla P+\frac{1}{c} {\bf{j}}
\times {\bf{B}}- m_{i} n \nabla \Phi,\\
\label{2}
\end{equation}
\begin{equation}
\frac{D  } {D t}(P n^{-\gamma})=0,\\
\label{3}
\end{equation}
\begin{equation}
\frac{\partial {\bf{B} }} {\partial t}+ c \nabla \times
 {\bf{E}}=0,\,\nabla \cdot {\bf{B}}=0,
\label{4}
\end{equation}
\begin{equation}
 {\bf {E} }=  -  \frac{1}{c}{\bf{V}} \times {\bf{B}} +
\frac{1}{ec} \frac{\bf{j}
  \times \bf{B} }{n_{e} },\,\,\,{\bf {j}} =
 \frac{c}{4 {\pi}}
 \nabla \times {\bf{B}}.\,\,\,\,
\label{5}
\end{equation}
Here the standard cylindrical coordinates   $\{r, \theta,z\}$
are adopted throughout the paper with the associated unit basic vectors
$\{  \bf{i}_r,\bf{i}_\theta, \bf{i}_z \}$; $\bf{V}$    is the plasma velocity;
$t$ is time; $D/Dt=\partial/\partial t+(\bf{V} \cdot\nabla)$  is the material derivative;
$\Phi(R)=-GM_c/R$ is the gravitational potential of the central object;  $R^2=r^2+z^2$;
$G$ is the gravitational constant; $M_c$ is the total mass of the central object.
The electric field  $\bf{E}$   is described by the generalized Ohm's law
which is derived from the momentum equation for the electrons fluid by neglecting the electrons inertia and pressure;
$\bf{B}$  is the magnetic field,  $\bf{j}$ current density;  $\gamma$  is the polytrophic coefficient,
($\gamma=5/3$ in the adiabatic case).  $P=P_e+P_i+P_n$ is the total plasma pressure;
$c$ is speed of light;  $P_l$  and  $m_l$ are the species pressures and masses ($l=e,i,n$ ),
subscripts $e$, $i$ and $n$ denote the electrons, ions and neutrals, respectively; $e$ is the electron charge;
$m_i=Zm_p$,  $m_p$ is the proton mass ($Z=1$ for simplicity).
Since the plasma is assumed to be quasi-neutral and partially ionized with small ionization degree  and strongly
coupled ions and neutrals
\begin{equation}
n_e \approx n_i \approx \alpha n_n , n \approx n_n, \alpha={n_e/(n_e+n_n)}<<1, V_i\sim V_n.
\label{6}
\end{equation}

In real protoplanetary disks, the electron density is determined by ionization versus recombination,
and the fractional ionization may vary significantly in space. In such cases, rate equations that
describe the ionization degree should be employed. However, in the present study the ionization
fraction is assumed to be constant. This simplification is made to avoid the widely uncertain
physics of ionization and recombination. Nevertheless, such approximation allows one to roughly
estimate the real characteristics of the equilibrium system (see a discussion in section 5
in \cite{Shtemler et al. 2007}).

\subsection{Scaling procedure}
The physical variables are now transformed
into nondimensional variables:
\begin{equation}
f_{nd}=f / f_*,
\label{7}
\end{equation}
where  $f$   and  $f_{nd}$ stand for any of the physical dimensional and
non-dimensional variables, while the characteristic scales $f_*$
are defined as follows:
\begin{equation}
V_*=\Omega_* r_* ,\,\,t_*=\frac{1}{\Omega_*},\,\,
\Phi_*={V_*}^2,\,\, m_*= m_i,\,\,
n_*=n_n,\,\,\,
 P_*=K(m_* n_*)^\gamma,\,\,
\,\,j_*=\frac{c}{4\pi}\frac{B_*}{r_*},\,\, E_*=\frac{V_*
B_*}{c}.
\label{8}
\end{equation}
Here $\Omega_*=(GM_c/r_*^3)^{1/2}$  is the Keplerian
angular velocity of the fluid at the characteristic radius $r_*$ that belongs to the Keplerian portion of the disk;
%
 %
 $K$ is the dimensional constant in the steady-state
polytropic law $P=Kn^\gamma$.
The characteristic values of the electric current and field, $j_*$  and  $E_*$, have been chosen consistently with Maxwell equations. The characteristic dimensional magnetic field  $B_*$    is specified below. %
Note that a preferred direction is tacitly defined here, namely, the positive direction of the $z$  axis is chosen according to positive Keplerian rotation.

The resulting dimensionless system (omitting the subscript $'nd'$ from the dimensionless variables) is given by:
\begin{equation}
\frac{\partial n} {\partial t}
               + \nabla \cdot(n {\bf{V}} ) =0,\,\,\,\,\,\,\,\,\,\,\,\,\,\,
\label{9}
\end{equation}
\begin{equation}
n \frac{ D \bf {V} }{Dt} = -\frac{1}{M_S^2}\nabla P +
\frac{1}{\beta M_S^2} {\bf{j}} \times {\bf{B}}-  n \nabla
\Phi,\,\,\,\Phi(r,z)=-{1 \over (r^2+z^2)^{1/2}},
\label{10}
\end{equation}
\begin{equation}
\frac{D (P n^{-\gamma})}{Dt}=0,
\label{11}
\end{equation}
\begin{equation}
\frac{\partial {\bf{B} }} {\partial t}+  \nabla \times
 {\bf{E}}=0,\,\nabla \cdot {\bf{B}}=0,
\label{12}
\end{equation}
\begin{equation}
 {\bf {E} }=  -  {\bf{V}} \times {\bf{B}} + \Pi_H \frac {\bf{j}
\times \bf{B} }{n},\,\,\,\,{\bf {j}} =
  \nabla \times {\bf{B}}.
\label{13}
\end{equation}
Here  $M_S$ and $\beta$ are the  Mach
number and plasma beta, and $\Pi_H$ is the Hall coefficient in the
generalized Ohm' law (13):
\begin{equation}
M_S=\frac{V_*}{c_{S*}},\,\,\,\beta=4\pi \frac{P_*}{B_*^2},\,\,\,
\Pi_H=\frac{\Omega_i}{ \Omega_*}\bigg{(}
\frac{l_i}{r_*}\bigg{)}^2\equiv \frac{B_* c}{4 \pi e \alpha n_*
\Omega_*r^2_*},
\label{14}
\end{equation}
$c_{S*}=\sqrt{P_*/(m_*n_*)}$ is the
characteristic  sound velocity, $l_i=c/\omega_{pi}$  and
$\Omega_{i}=eB_*/(m_* c)$ are the inertial length and the Larmor
frequency of  ions, respectively,
$\omega_{pi}=\sqrt{4\pi e^2 \alpha n_{*}/m_*}$ is the plasma frequency
of the ions. The increasing importance of the Hall term for weakly ionized disks is
apparent as  $\Pi_H$ 
is inversely proportional to the ionization degree $\alpha$.

A common property of thin Keplerian disks is their highly
compressible motion with large Mach numbers $M_S$.
Furthermore, the characteristic effective semi-thickness  $H_*=H(r_*)$ of the disk ($H=H(r)$  is the local disk height) is such that the disk aspect ratio $\epsilon$  equals the inverse Mach number:
\begin{equation}
\frac{1}{M_S}=\epsilon=\frac{H_*}{r_*}\lesssim1.
\label{15}
\end{equation}

The smallness of  $\epsilon$   means that dimensional axial coordinate  $|z|\leq H$ is also small, i.e.  $z/r_*\sim \epsilon $, and the following rescaled values of the order of unity in  $\epsilon$   may be introduced in order to further apply the asymptotic expansions in   $\epsilon$   (Shtemler et al 2007):
\begin{equation}
\zeta=\frac{z}{\epsilon},\,\,\,h =\frac{H}{\epsilon},\,\,\,\pi_H =\frac{\Pi_H}{\epsilon},
\label{16}
\end{equation}
where as was mentioned in the introduction, the rescaled Hall parameter $\pi_H $
is determined in terms of the characteristic disk thickness $H_*=\epsilon r_*$  as the density inhomogeneity length.

\subsection{Physical conditions in protoplanetary disks}
Before turning to the detailed description of the Hall equilibrium
 of rotating Keplerian disks it is instructive to estimate some of
 the parameters that have been introduced in the previous subsection.
 Typical protoplanetary disks extend up to the order of $100 AU$
 and consist of molecular gas with characteristic ion masses between
 $30m_p$ to $40m_p$. The temperature within a radius of about $0.1 AU$ may exceed $10^3 K$
 were as further away from the central star the temperature decreases and may reach
 values of about $10 K$ in the outer regions of the disk.
 As a result, while in the inner regions of the disk thermal collisions provide
 the dominant ionization mechanism, in regions beyond about $0.1 AU$ the only sources of ionization are nonthermal,
 like cosmic rays and the decay of radioactive elements (\cite{Umebayashi and Nakano 1988}, \cite{Gammie 1996},
  \cite{Igea and Glassgold 1999}).
  Taking that into account, the frequently used model is employed, in which the radial temperature profile is given by
\begin{equation}
T(r)=280\large (\frac{r}{1AU} \large ) ^{-1/2}\;\;\;\;K,
\label{eq:temp}
\end{equation}
while the column mass density is given by the minimum mass model as
\begin{equation}
\Sigma(r)=1700\large (\frac{r}{1AU} \large ) ^{-3/2} \;\;\;\;g\;cm^{-2}.
\label{eq:mass}
\end{equation}
The thickness of the disk as well as the neutral number density radial profiles may be calculated from the two functions above. The ionization fraction is given by
\begin{equation}
\alpha = \sqrt {\frac{\xi }{n\beta_r }}, \label{eq:ionization}
\end{equation}
where $\xi$, the ionization rate due to the non thermal processes, is given by
\begin{equation}
\xi = 10^{-17}exp(-\Sigma /192)+7\times 10^{-23} \;\;\;\;sec^{-1},
\label{eq:rate}
\end{equation}
and $\beta_r$ is the dissociative recombination rate given by
\begin{equation}
\beta_r = 1.1\times 10^{-7} \large ( \frac{T}{300 K} \large )
^{-1}\;\;\;\;cm^3\;sec^{-1}. \label{eq:recombination}
\end{equation}
A detailed and critical study of the model as well as some other
suggestions may be found in \cite{Hayashi et al 1985},
\cite{Gammie 1996}, \cite{Fromang et al. 2002}, \cite{Sano and
Stone 2002}. Here however, Eqs.
(\ref{eq:temp})-(\ref{eq:recombination}) are employed in order to
obtain a rough estimate for the physical conditions in the disk.
To do that, attention is focused on two representative radii, $r_1
=1AU$ and $r_2 =20AU$. A characteristic magnetic field of
$0.1G$ is assumed while the mass of the central star has been assumed to be one solar mass.
The results are summarized in Table (1). The
increase in the rescaled Hall parameter is due to the decrease in
the neutrals number density that compensates for the increase in
the ionization fraction and the angular momentum per unit mass. In
particular, it is evident that the entire disk resides within the
Hall regime.

\begin{table*}
 \centering
 \begin{minipage}{140mm}
\caption{Estimation of the physical conditions at two radii within the disk}
\begin{tabular}{@{}ccccccc@{}}
\hline $~~r_*\; AU~~$ & $~~~~\alpha~~$ & $~~~l_i\;Km~~$
& $~~~\pi_H~~$ & $~~~\omega_{pi}\;sec^{-1}~~$ & $~~~\Omega_*\;sec^{-1}~~$ &
\\
\hline
1  & ~~$1.12\times 10^{-14}$  & $1.24\times 10^{3}$  & 1.03  & 241.6 & $2.00\times 10^{-7}$  \\
20  & ~$4.14\times 10^{-13}$  & $1.25\times 10^{4}$ & 11.2 & 23.82 & $2.23\times 10^{-9}$ \\
\hline
\end{tabular}
\end{minipage}
\end{table*}

\section{HALL EQUILIBRIUM IN THIN DISKS IN 3D AXISYMMETRICAL MAGNETIC FIELDS}
In magnetized disks the solution of the steady-state equilibrium problem ($\partial/\partial t \equiv0$)
may be obtained within the Keplerian portion of a disk by asymptotic expansions in small $\epsilon$
 with the aid of Eqs.
(\ref{9})-(\ref{16})
  (similar to \cite{Regev 1983},  \cite{Ogilvie 1997},  \cite{Kluzniak and Kita 2000},
   \cite{Umurhan et al. 2006}, and \cite{Shtemler et al. 2007}).
   This yields for the gravitational potential in the Keplerian portion of the disk:
\begin{equation}
\Phi(r,\zeta)=-\frac{1}{r}+\frac{\epsilon^2\zeta^2}{2r^3}+O(\frac{\epsilon^4}{r^5}),\,\,\,
 \epsilon\ll r,\,\,\zeta\sim\epsilon^0.
\label{22}
\end{equation}

To leading order in $\epsilon$  within the thin disk approximation, the toroidal velocity   $V_{\theta}$ is described by the Keplerian law, which follows from the leading order radial component of the momentum equation:
\begin{equation}
{\bf{V}}=V_\theta {\bf{i}}_\theta ,\,\,\,
V_{\theta}=\Omega(r)r,\,\,\, \Omega(r)=r^{-3/2},\,\,\,V_\theta\sim \epsilon^0,\,\,\,V_r\sim \epsilon,\,\,\,V_z\sim \epsilon^2.
\label{23}
\end{equation}
The asymptotic orders in   $\epsilon$  in equilibrium relations
(\ref{23})   are similar to those in \cite{Ogilvie 1997}. They may
be inferred by observing that due to Eq. (\ref{9}) ($\nabla \cdot
(n{\bf{V}})=0$) the axisymmetric fluid momentum per unit volume is
expressed in terms of  a scalar stream function $\chi(r,\zeta)$
\begin{equation}
n{\bf{V}}=nV_\theta {\bf{i}}_\theta +\frac{1}{r}\hat{\nabla}\chi \times {\bf{i}}_\theta,\,\,\,
\big{(}\hat{\nabla}=
{\bf{i}}_r \frac{\partial}{\partial r}+{\bf{i}}_z \frac{1}{\epsilon}\frac{\partial}{\partial \zeta}\big{)}.
\label{24}
\end{equation}
Assuming now that the toroidal velocity component is dominant, and
noticing that   $\partial/\partial z\sim \epsilon^{-1}$ imply that
$ \chi \sim \epsilon^2$, which immediately results in the ordering
presented in Eq. (\ref{23}) (see also discussion just after Eq.
(\ref{41}) ).

By a similar way   the divergent free axisymmetric magnetic field $\bf{B}$, ($\nabla \cdot  \bf{B}=0$)
 is written in terms of a poloidal flux function $\Psi(r,\zeta)$ in the following way:
\begin{equation}
{\bf{B}}=B_\theta {\bf{i}}_\theta +\frac{1}{r}\hat{\nabla}\Psi \times {\bf{i}}_\theta,\,\,\,
\Psi(r,\zeta)= \Psi_0(r)+\epsilon  \psi(r,\zeta),\,\,\,B_r=-\frac{1}{r}\frac{\partial \psi}{\partial \zeta},
\,\,\,B_z=\frac{1}{r}\frac{d\Psi_0}{dr}+\epsilon \frac{1}{r}\frac{\partial\psi}{\partial r}.
\label{25}
\end{equation}
Here the poloidal flux function  $\Psi$ is presented as sum of
$\Psi_0(r)$ and  $\epsilon\psi(r,\zeta)$, which are scaled in  $\epsilon$ by such a way to produce
   $B_z\sim B_r\sim\epsilon^0$ to leading order in the thin disk approximation.

Proceeding further it is  noticed that Eq. (\ref{12}) means that
the electric field is derived from a potential  $\phi(r,\zeta)$
\begin{equation}
{\bf{E}}=\epsilon \hat{\nabla}\phi.
\label{26}
\end{equation}
Taking now the dot product of both sides of Eq. (\ref{13}) with
the magnetic field reveals that
${\bf{B}}\cdot\hat{\nabla}\phi=0$, and therefore $\phi$ is a
function of the magnetic flux, i.e.
\begin{equation}
\phi=\phi(\Psi),
\label{27}
\end{equation}
which immediately, due to axisymmetry means that the toroidal component of the electric field vanishes,
i.e.,  $E_\theta=0$. In order to impose that condition, Ampere's law is first employed in order to write
the electric current density in the following way:
\begin{equation}
{\bf{j}}=j_\theta{\bf{i}}_\theta
+\frac{1}{r}\hat{\nabla}(rB_\theta)
\times{\bf{i}}_\theta,\,\,\,
  j_r=-\frac{1}{\epsilon}\frac{\partial B_\theta}{\partial \zeta},\,\,\,
j_\theta=-\frac{1}{r}\frac{d}{d r}\big{(}\frac{d\Psi_0}{d r}\big{)}
-\frac{1}{\epsilon}\frac{1}{r}\frac{\partial^2 \psi}{\partial \zeta^2}
-\epsilon \frac{\partial }{\partial r}\big{(}\frac{1}{r}\frac{\partial \psi }{\partial r}\big{)}
,\,\,\,j_z=\frac{1}{r}\frac{\partial (rB_\theta)}{\partial r}.
\label{28}
\end{equation}
Returning now to Eq. (\ref{13}), the toroidal component of the
electric field is given by:
\begin{equation}
E_\theta={\bf{B}}\cdot\hat{\nabla}(rB_\theta)=0.
\label{29}
\end{equation}
The leading order in  $\epsilon$ of the last equation yields
\begin{equation}
\frac{\partial B_\theta}{\partial \zeta}\frac{d\Psi_0}{dr}=0,
\label{30}
\end{equation}
which is satisfied if either  $\partial B_\theta/\partial \zeta=0$ or  $d\Psi_0/dr=0$.
The former case was considered by \cite{Ogilvie 1997} within the classical MHD
model for the equilibrium of a differentially rotating thin disk containing a pure poloidal
magnetic field ($B_\theta\equiv0$). The second case, for the pure toroidal magnetic field ($\Psi\equiv0$)
has been investigated within the Hall MHD model by \cite{Shtemler et al. 2007}. Thus in the leading order
in  $\epsilon$, there are three possible equilibrium states: (i) pure toroidal, (ii) pure poloidal and (iii)
mixed magnetic field equilibrium, which can not be reduced from one to the other. In the present work
the case  $\Psi_0\equiv0$  is investigated within the Hall MHD model allowing for both toroidal
as well as poloidal components of the magnetic field in the leading order. This results in:
\begin{equation}
{\bf{B}}=B_\theta {\bf{i}}_\theta +\frac{1}{r}\hat{\nabla}\Psi
\times {\bf{i}}_\theta,\,\,\, \Psi(r,\zeta)= \epsilon
\psi(r,\zeta),\,\,\,
 \phi=  \phi(\psi),\,\,\,\,
B_r=-\frac{1}{r}\frac{\partial\psi}{\partial \zeta},
\,\,\,\hat{B}_z\equiv\frac{B_z}{\epsilon
}=\frac{1}{r}\frac{\partial\psi}{\partial r},
\label{31} \end{equation}
 where  the magnetic flux function $\Psi$
is scaled with  $\epsilon$  in such a way that the resulting
radial component of the magnetic field is of the order of the
toroidal one ($\sim\epsilon^0$) in the thin disk approximation,
while the axial magnetic field  $B_z$ is of order  $\sim\epsilon$.
Consequently Eqs. (\ref{29}) and (\ref{31}) yield
\begin{equation}
 rB_\theta=I(\psi),
\label{32}
 \end{equation}
 where according to Ampere's law  $I(\psi)$ is the current flowing through a circular area in the plane  $\zeta=const$.
  Furthermore, now Ampere's law may be written as follows:
 \begin{equation}
 {\bf{E}}\equiv\epsilon\hat{\nabla}\phi=-{\bf{V}}\times {\bf{B}}+\epsilon \pi_H
 \frac{ {\bf{j}}\times {\bf{B}}}{n}.
\label{33}
 \end{equation}

Finally  note that the case of the pure toroidal magnetic field
$\psi\equiv0$ (i.e. ${\bf{B}}=(0,B_\theta,0$)) requires a separate consideration.
In that case the relation $E_\theta=0$ is satisfied identically, while
Faraday's law reduces to with the help to Eq. (\ref{33}) to
\begin{equation}
 \hat{\nabla}\times\big{[} \frac{rB_\theta}{N}\hat{\nabla}(rB_\theta)\big{]}=0,
\label{34}
 \end{equation}
where  $N=r^2n$ is the inertial moment density through the Keplerian disk.
This leads to the familiar result up to the scaling factor  $\pi_H$ (\cite{Shtemler et al. 2007}):
\begin{equation}
 rB_\theta=I(N),\,\,\,
 \pi_H\phi=-\int_0^N\frac{ I(N) \dot{I}(N)}{N}dN.
\label{35}
 \end{equation}
Here and everywhere below the upper dot denotes a derivative with
respect to the argument.

Every thing is ready now to derive the relevant GS equation in the
general case $\psi\neq0$. The latter is indeed readily obtained by
projecting Eq. (\ref{13}) onto the  $\hat{\nabla}\psi$ direction
and taking into account the above relations. To leading order in
$\epsilon$ the result is the following nonlinear differential
equation for  $\psi$, which depends parametrically on the radial
coordinate  $r$:
\begin{equation}
\frac{\partial^2\psi}{\partial\zeta^2}=-I(\psi)\dot{I}(\psi)- \frac{1}{\pi_H}N(r,\zeta)[\Omega(r)+\dot{\phi}(\psi)].
\label{36}
 \end{equation}
This is the GS equation (similar to \cite{Lovelace et al. 1986},
\cite{McClements and Thyagaraja 2001} and references therein) that
is modified for the special case of a thin disk in Hall MHD
equilibrium. It interesting to note that in the limit of
negligible Hall effect ($\pi_H\to 0$) Eq. (\ref{36}) is reduced to
the algebraic limit of the familiar isorotation law of ideal MHD
that expresses the fluid's angular momentum constancy on magnetic
surfaces.

In general, a coupled problem should be considered inside and outside the Keplerian disk such that the edge
of the disk  $\zeta=h(r)$ is determined self-consistently. However, to avoid the solution of the outer problem,
the following assumption that is supported by the observation data for some protoplanetary disks
(\cite{Calvet et al. 2002}) is adopted:
\begin{equation}
h(r)\equiv const=1.
\label{37}
 \end{equation}
Furthermore, the differential equation (\ref{36}) must be
complemented by an appropriate boundary conditions at the disk
edge $\zeta=h(r)$, as well as by symmetry condition at the
midplane (assuming  $\hat{B}_z$, as well as  $B_\theta$ being even
functions of  $\zeta$):
\begin{equation}
  n=0,  \psi=\Gamma(r) \,\,\,\, \mbox{at}\,\,\,\,    \zeta=h(r), \,\,\,\,
 \frac{\partial \psi }{\partial \zeta }=0\,\,\,\,   \mbox{at}\,\,\,\,   \zeta=0,
\label{38}
 \end{equation}
where  $\Gamma(r)$ is a specified function. Since there is no
direct observational information on how magnetic fields are
distributed on disk edges the following specific equilibrium
configurations are conjectured which are idealized representations
of Keplerian disks. The value of  $\hat{B}_z$ at the horizontal
disk edges is assumed to be a sum of a constant primordial
(\cite{Ogilvie 1997}) and a dipole-like (\cite{Lepeltier and Aly
1996}, see  also \cite{Lovelace et al. 2002}, and \cite{Matt et
al. 2002}) magnetic fields. The latter qualitatively reflects the
effect of the central body on Keplerian portion of the disk.
Below the two following configurations  are examined:
\begin{equation}
 \Gamma(r)=0.5Br^2-Mr^{-1},\,\,\,\,     ( B=\pm 1),
\label{39}
 \end{equation}
\begin{equation}
\Gamma(r)= -Mr^{-1},\,\,\,\,        ( B=0,\,\,M=\pm 1),
\label{40}
 \end{equation}
where the constants  $B$ and  $M$ are the dimensionless primordial
magnetic field strength, and the moment of the dipole-like
magnetic field, respectively. In the case (\ref{39}) the
characteristic magnetic field  $B_*$ is chosen to be equal to the
primordial magnetic field, while in the case (\ref{40}) of zero
primordial magnetic field,  $B_*$  is chosen to make the
dimensionless magnetic moment equal to unity. The signs of  $B$
and  $M$  depend on their directions with respect to positive
direction determined by Keplerian rotation. Both  $B$ and  $M$
correspond to the magnetic flux function scaled by  $\epsilon$, so
the corresponding unscaled values are small constant of the order
of  $\epsilon$. Since the primordial magnetic field is likely much
smaller than that of the dipole like magnetic field, the mixed
case (\ref{39}) with  $B\sim M$ is rather a qualitative
demonstration of the method ability to describe a wide variety of
structures, while the case (\ref{40}) assumes implicitly that the
primordial magnetic field creates a much smaller input into the
boundary conditions than the magnetic dipole. Although, in general
the magnetic axis of the dipole is not aligned with the disk axis,
some of the important problems, may be investigated analytically
in an axisymmetric approach, which is valuable for subsequent full
3D analysis.

Note that  Ohm's law  generalized for Hall equilibrium allows to
integrate explicitly the vertical momentum balance equation in
thin disk approximation. However, this procedure is singular in
the limit of small Hall parameter. Employing Eqs. (\ref{33}) and
(\ref{23}) in the axial component of the momentum equation (10)
provides an explicit relation for the number density to the
leading order in $\epsilon$:
\begin{equation}
n(r,\zeta)= \nu
\big{[} \frac{2}{\beta\pi_H}\big{(} \phi(\psi)-\phi(\Gamma)\big{)}
+\frac{2\Omega}{\beta\pi_H}\big{(} \psi-\Gamma\big{)}
+\frac{1-\zeta^2}{r^3}\big{]}^{\frac{1}{\gamma-1}},\,\,\,\,\,N=r^2n,\,\,\,\,\,
 \nu=\big{(}\frac{\gamma-1}{2\gamma}\big{)}^{\frac{1}{\gamma-1}}.
\label{41}
 \end{equation}
Thus to the leading order, the momentum equation in the axial
direction  yields the density distribution (\ref{41}), the
momentum equation in the radial direction   provides  the
Keplerian toroidal velocity $\sim\epsilon^0$, while the momentum
equilibrium in the azimuthal direction along with the mass balance
relation determine the radial and axial velocities to be small of
the order of  $\epsilon$ and  $\epsilon^2$, respectively.
Expression (\ref{41}) is singular for vanishing  $\pi_H$ and
everywhere below the Hall parameter $\pi_H$,  is assumed to be of
the order of or much larger than unity.

Inspecting the resulting problem
 (\ref{36}) and (\ref{41}) reveals that it contains the Hall parameter
 $\pi_H$ in the two following combinations: as the ratio with inertial moment density
 $N/\pi_H$ in Eq. (\ref{36}), and as a factor with the plasma beta  $\beta\pi_H$ in
 Eq. (\ref{41}).
 Since both plasma beta and Hall parameter are independent physical parameters,
 and since they are generally unknown in astrophysical systems, a scaled Hall plasma beta parameter
 is naturally introduced:
\begin{equation}
 \beta _H=\beta\pi_H.
\label{42}
 \end{equation}
Note also that for large Hall plasma beta  $\beta_H$, the
arbitrary function  $\phi(\psi)$ is dropped from Eqs. (\ref{41})
and the number density is approaching its pure hydrodynamical
limit.

The arbitrary functions  $\phi(\psi)$  and  $I(\psi)$  reflect the uncertainty of the steady-state equilibrium solution that as assumed here describes the equilibrium solution at long times. In general the corresponding initial value problem must be solved which shifts that uncertainty to the arbitrariness of the initial data (see e.g. Lovelace et al. 2002). The magnetic field and density structure is governed by the choice of the functions $\phi(\psi)$  and  $I(\psi)$, here taken as following trial functions in order to qualitatively describe the equilibrium states:
\begin{equation}
 I(\psi)=I^{(0)}+I^{(1)}\psi,\,\,\phi(\psi)=\phi^{(0)}+\phi^{(1)}\psi.
\label{43}
 \end{equation}
The constant  $ I^{(0)}$ is the total current through the disk
that is localized along the axis (\cite{Shtemler et al. 2007}),
while  $ \phi^{(0)}=0$ without loss of generality due to the
physical meaning of the electric field potential (see also Eqs.
(\ref{36}) and (\ref{41})). Such kind simplest trial function are
commonly used for  ideal MHD model (see e.g. \cite{Wu and Chen
1989}, \cite{Kaiser and Lortz 1995}, \cite{Bogoyavlenskij 2000}
and references therein for astrophysical applications). Sometimes
the above relations are considered as first terms of generic
expressions in which the arbitrary functions can be expanded in
power series in  $\psi$ (\cite{Bogoyavlenskij 2000}). Clearly,
approximations (\ref{43}) are biased by their partial form.
However, in the present study it is conjectured that the main
features of the equilibrium are rather general since relations
(\ref{43}) may be considered as a best fit linear approximations
rather than first terms in appropriate Taylor' sets.

To summarize, in the present model Hall MHD axisymmetric equilibria are determined
up to the Hall  parameter  $\pi _H$, the Hall plasma beta  parameter  $\beta _H$, and the free constants
$B$ and  $M$ that characterize the primordial magnetic field and an effective magnetic dipole moment induced
by the central body. In such equilibria the electric potential as well as the toroidal magnetic field
are arbitrary functions of the magnetic flux, while the density is an explicit function of
the coordinates and the magnetic  flux function which satisfies a GS type equation  modified for the
special case of  Hall equilibrium in thin-disk approximation.

\section{ANALYSIS OF THE EQUILIBRIUM IN TERMS OF the scaled INVERSE HALL PARAMETER}
Inspecting expression (\ref{41}) for the inertial moment density
$N$ and the GS equation (\ref{36}) for the magnetic flux   $\psi$
that contains  $N/\pi_H$  reveals the existence of the following
small parameter  $\delta$:
 \begin{equation}
 \delta=\frac{\nu}{\pi_H}\lesssim1.\,\,\,
%
\label{44}
  \end{equation}
 In the adiabatic case with  $\gamma=5/3$ adopted everywhere below, this
  parameter is indeed sufficiently small within the range  from large values
  $\pi_H$   up to   $\pi_H=1$   when it equals  $\delta=\nu\approx 0.1$ (see definition  of $\nu$ in Eq. (36)).
  Note also that validity for using  $\delta=\nu/\pi_H$ as the small parameter is violated in the classical MHD limit
   ($\pi_H\to 0$)  which requires a separate consideration.

Since the parameters of the equilibrium are widely arbitrary, two
different families of  equilibrium solutions are found, which
correspond to either (i) small ( $\sim\delta$) or (ii) finite (
$\sim\delta^0$) deviations of the magnetic flux $\psi(r,\zeta)$
from its boundary value  $\Gamma\sim\delta^0$. Note that the
scaling in  $\delta$ is applied to the flux function that is
already scaled in  $\epsilon$ (see Eqs. (\ref{25})), so that the
dimensionless unscaled value   $|\psi-\Gamma|$ is either (i)
$\sim\epsilon \delta$ or (ii)  $\sim\epsilon \delta^0$.
Consequently, as will be seen below, the Hall equilibrium disks fall into two types which characterized by quite different orders in $\delta$: Keplerian disks   with (i) small ($R_d\sim\delta^0$) and (ii) large (${R_d}\gtrsim\delta^{-k}$, $k>0$) radius of the disk.
%
Thereafter the two families of the equilibrium solutions are named, respectively,
Keplerian disks with small- and  large- radius.
In both families the angular velocity satisfies  the Keplerian law in the leading order in
$\epsilon$  independently of  $\delta$.
Thus, using the small parameter  $\delta$  not only enables and simplifies further calculations,
but also allows to clearly distinguish two quite different Hall
equilibrium families of solutions that are considered below separately.

\subsection{First family of solutions: small-radius  disks}
Assuming that $|\psi-\Gamma|\sim\delta^0$ the solution is expanded in power series in  $\delta$:
 %
  \begin{equation}
  \psi(r,\zeta)=\psi_0+\delta\psi_1+\dots,\,\,
    N(r,\zeta)/\nu=N _0+\delta N_1+\dots,\,\, n(r,\zeta)/\nu=n _0+\delta
    n_1+\dots.
 I(\psi)=I(\psi_0)+\dots,\,\, \phi(\psi)=\phi(\psi_0)+\dots.
\label{45}
  \end{equation}
Substituting Eqs. (\ref{45}) into (\ref{36}) - (\ref{41}) yields
to leading order in  $\delta$ (omitting the subscript 0 for
brevity)
  \begin{equation}
  \frac{\partial^2\psi}{\partial\zeta^2}=-I(\psi)\dot{I}(\psi),\,\,
\psi=\Gamma(r)\,\, \mbox{at}\,\,\zeta=1,\,\,
\frac{\partial \psi}{\partial \zeta }=0\,\, \mbox{at}\,\,\zeta=0.
\label{46}
 \end{equation}

Thus, in the lowest order in  $\delta$ the electric field potential is dropped out from
 the equilibrium problem. Transforming then to a new independent variable
 $Y=\partial\psi/\partial\zeta$ and integrating the transformed differential equation
 in (\ref{46})
 under the appropriate conditions at  $\zeta=0$  yields
  \begin{equation}
  \frac{\partial\psi}{\partial\zeta}=\sqrt{I^2(\psi(r,0))- I^2(\psi(r,\zeta))}.\,\,
\label{47}
 \end{equation}
 Integrating Eq. (\ref{47}) once more under the conditions at  $\zeta=1$ in (\ref{46})
  yields to leading order in  $\delta$ the following implicit dependence of  $\psi$
  and   $N=r^2n$ on  $r$ and  $\zeta $:
 \begin{equation}
1-\zeta=-\int_{\Gamma(r)}^{\psi(r,\zeta)}\frac{d\psi}{\sqrt{I^2(\Gamma)- I^2(\psi)}},\,
\label{48}
  \end{equation}
  and
  \begin{equation}
  N(r,\zeta)=r^2n(r,\zeta)=r^2\big{[}\frac{2}{\beta_H }\big{(} \phi(\psi)  -\phi(\Gamma)    \big{)}
 +\frac{2\Omega(r)}{\beta_H}\big{(} \psi - \Gamma\big{)}
 +\frac{1-\zeta^2}{r^3}
 \big{]}^{3/2}.
\label{49}
  \end{equation}
Thus both poloidal and toroidal magnetic fields  are completely
characterized by  arbitrary functions  $I(\psi)$,  $\phi(\psi)$
and the edge magnetic flux function $\Gamma(r)$ (with the
characteristic amplitudes $B$ and $M$). The function  $N(r,\zeta)$
depends also on  $\phi(\psi)$, $\Gamma(r)$,  and on the Hall
plasma beta  $\beta_H$.

Substituting (\ref{43}) into Eqs. (\ref{48})-(\ref{49}) yields for
$\gamma=5/3$:
  \begin{equation}
\psi= \Gamma(r) +\big{(}\frac{I^{(0)}} {I^{(1)}} +\Gamma(r) \big{)}
 \big{(}\frac{cos(I^{(1)}\zeta)} { cos(I^{(1)})} -1 \big{)},
\label{50}
  \end{equation}
 \begin{equation}
n=  \big{[}
\frac{2}{\beta_H }
 \big{(}\frac{I^{(0)}} {I^{(1)}} +\Gamma(r) \big{)}
\big{(}\phi^{(1)} +\Omega(r) \big{)}
\big{(}\frac{cos(I^{(1)}\zeta)} { cos(I^{(1)})} -1 \big{)}
+\frac{1-\zeta^2}{r^3}
\big{]}^{3/2},\,
\label{51}
  \end{equation}
  \begin{equation}
 rB_{r}=\big{(}I^{(0)}+I^{(1)}\Gamma(r) \big{)}\frac{sin(I^{(1)}\zeta)} { cos(I^{(1)})},\,\,
rB_{\theta}=\big{(}I^{(0)}+I^{(1)}\Gamma(r)
\big{)}\frac{cos(I^{(1)}\zeta)} { cos (I^{(1)})},\,\,
B_{z}=\frac{1}{r}\frac{d\Gamma}{dr}\frac{cos(I^{(1)}\zeta)} { cos
(I^{(1)})}.
\label{52}
 \end{equation}
 In particular, in the midplane   $\zeta =0$ Eqs. (\ref{50})-(\ref{52}) results in:
  \begin{equation}
 n=  \big{[}
\frac{2}{\beta_H }
 \big{(}\frac{I^{(0)}} {I^{(1)}} +\Gamma(r) \big{)}
\big{(}\phi^{(1)} +\Omega(r) \big{)}
\frac{1-cos(I^{(1)})} { cos(I^{(1)})}
+\frac{1}{r^3}
\big{]}^{3/2},\,
  B_{r}=0,\,\,
rB_{\theta}=\frac{I^{(0)}+I^{(1)}\Gamma(r)} { cos (I^{(1)})},\,\,
B_{z}=\frac{1}{r}\frac{d\Gamma}{dr}\frac{1} { cos (I^{(1)})},
\label{53}
  \end{equation}
 while on the disk's edges   $\zeta =\pm 1$ the magnetic field and number density are given by:
 \begin{equation}
 n= 0,\,\,
rB_{r}=\pm\big{(}\frac{I^{(0)}} {I^{(1)}} + \Gamma(r) \big{)}tan (I^{(1)}),\,\,
rB_{\theta}=I^{(0)}+I^{(1)} \Gamma(r),\,\,
B_{z}=\frac{1}{r}\frac{d\Gamma}{dr}
\label{54}
  \end{equation}

The parameters  $I^{(0)}$,  $I^{(1)}$,  $\phi^{(1)}$,   $\beta_H$ and  $B$ and/or
  $M$ are need for a complete description of the Hall equilibrium state.
  Furthermore, the solutions (\ref{50})-(\ref{51}) are invariant with
  respect to simultaneous change of signs of $I^{(0)}$ and  $I^{(1)}$,
  so further analysis is restricted to positive values of  $I^{(0)}$.

Inspecting Eq. (\ref{51}) it is readily seen that a class of
equilibria may be found which is characterized by a finite radius
that appears as a cut-off value at which the plasma density
vanishes. The finite radius of the disks,  $R_d$, may be inferred
from the condition of zero number density that according to Eq.
(\ref{51}) is given by:
 \begin{equation}
\frac{2}{\beta_H }
\big{[}\frac{I^{(0)}} {I^{(1)}} + \Gamma(R_d) \big{]}
\big{[}\phi^{(1)} +\Omega(R_d) \big{]}
\big{[}\frac{cos(I^{(1)}\zeta)} { cos (I^{(1)})}-1\big{]}
 +\frac{1-\zeta^2}{R_d^3}=0.
\label{55}
    \end{equation}
Note that the frequently used limit of large plasma beta reduces
Eq. (\ref{55}) to the classical hydrodynamic limit and yields
$R_{d}=\infty$.
Generally the parameters $I^{(0)}$, $I^{(1)}$, $\phi^{(1)}$,
$\beta_H$ and  $B$ and/or  $M$ may be correlated by using Eq.
(\ref{55}) with a given, e.g. observed, value of the disk radius.
Neglecting the term  $(1-\zeta^2)/R_d^3$ in Eq. (\ref{55}) (that
is a plausible assumption if the dimensionless disk radius is
sufficiently large, although corresponding values of $R_d$ will be
larger for larger values of the Hall plasma beta $\beta_H$), the
value   $R_d$ may be explicitly estimated as the root of the
following algebraic equation that is independent on the Hall
plasma beta:
 \begin{equation}
\big{[}\frac{I^{(0)}} {I^{(1)}} + \Gamma(R_d) \big{]}
\big{[}\phi^{(1)} +\Omega(R_d) \big{]}
\approx 0.
\label{56}
   \end{equation}
%
 Equation (\ref{56}) is satisfied by either its first or second co-factors  are equal to zero
 \begin{equation}
 R_{d1}^3+2\frac{I^{(0)}} {BI^{(1)}} R_{d1}-2\frac{M}{B}=0,
\label{57}
  \end{equation}
\begin{equation}
 R_{d2}= \big{(}- \frac{1}{\phi^{(1)}} \big{)}^{2/3},\,\,\, (\phi^{(1)}<0),
\label{58}
  \end{equation}
where the first root  $R_{d1}$ results also in zero values of both
$B_{r0}$ and  $B_{\theta0}$  at the lateral disk edge.
As follows from Eq. (\ref{55})
 the first root is governed by the effect of both toroidal and poloidal magnetic fields,
 while the second root is determined by the combined effect of rotation and poloidal magnetic field. %

In two limiting cases of predominantly either permodial ($M=0$) or
dipole ($B=0$) origin, the cubic equation Eq. (\ref{57}) has the
following simple explicit solutions, respectively
\begin{equation}
R_{d1}^{(B)}\approx\big{(}-\frac{2I^{(0)}}{BI^{(1)}}\big{)}^{1/2},
\,\,\, \big{(}\frac{I^{(0)}}{B I^{(1)}}<0\big{)},\,\,
\label{59}
  \end{equation}
\begin{equation}
R_{d1}^{(M)}=\frac{MI^{(1)}}{I^{(0)}},
\,\,\, \big{(} \frac{MI^{(1)}}{I^{(0)}}>0\big{)},
\label{60}
  \end{equation}
where   $R_{d1}^{(B)}\to \infty$  as  $I^{(1)}\to 0$,  and $R_{d1}^{(M)}\to \infty$ as  $I^{(0)}\to 0$.
Another interesting special solution of Eq. (\ref{57}) valid for
mixed magnetic flux at the horizontal flux, and sufficiently low
total currents $I^{(0)}$ is
   \begin{equation}
  R_d\approx  \root 3 \of {\frac{2M}{B}},\,\,\, \, \big{(}\frac{2M}{B}>0, \,\,\, \,\vert \frac{I^{(0)} }{ I^{(1)}}\vert<<
  \root 3 \of {\frac{M^2\vert B\vert}{2}}\big{)}
\label{61}
    \end{equation}
which means that $\Gamma(R_d)\approx0$. This corresponds  to zero
net poloidal flux, configurations that have been recently employed
in order to study the viability of internal dynamo action
(\cite{Brandenburg et al. 1995}, \cite{Fromang and Papaloizou
2007}). It should finally be noticed that the
finite radius solutions  obtained above are the direct result of the requirement
of finite disk thickness in problem (\ref{36})-(\ref{38})
considered for the special boundary function $\Gamma(r)$ in Eqs.
(\ref{39})-(\ref{40}) and the unique combinations of the
functional dependencies of the total current and electric
potential  on the magnetic flux function in Eq. (\ref{43}).
Alternatively, requiring a vertically diffused disk with exponentially decreasing
density (as is done in \cite{Coppi and Keyes 2003} and
\cite{Liverts and Mond 2008}) will result also in exponentially
decreasing radial density profiles.
%

\subsection{Second family of solutions: large-radius  disks}
For small deviations of the magnetic flux function from its edge
value, the above asymptotic expansions (in  $\delta$) fail. Thus,
assuming that the deviation of magnetic flux from its boundary
value is of order  $\delta$, it may be written:
 $I^{(0)}=\delta\bar{I}^{(0)},
\Gamma=\delta\bar{\Gamma}, (B=\delta\bar{B},
  M=\delta\bar{M}$) are scaled in $\delta$, and omitting
 the bars  with no confusion
%
%
\begin{equation}
\psi(r,\zeta) -\delta\Gamma(r)=\delta\psi_1+\dots,\,\,
N(r,\zeta)/\nu=N _0+\delta N_1+\dots,\,\,n(r,\zeta)/\nu=n
_0+\delta n_1+\dots, \,\,I(\psi)=\delta I^{(0)}+I^{(1)}
\delta\Gamma. \label{62}
  \end{equation}

Substituting (\ref{62}) into (\ref{36})-(\ref{41}) yields in the
leading order in $\delta$ for density and magnetic flux
 \begin{equation}
n_0=\big{(}
\frac{1-\zeta^2}{r^3}
\big{)}^{3/2},\,\,\,N_0=r^2n_0,
\label{63}
   \end{equation}
\begin{equation}
\frac{\partial^2\psi_1}{\partial\zeta^2}=-(I^{(0)}+I^{(1)}\Gamma)I^{(1)}
- N_0(r,\zeta)[\Omega(r)+\phi^{(1)}],\,\,
\psi_1 =0\,\,\,\,\mbox{for}\,\,\,\,\,\,\zeta=1,\,\,\,
\frac{\partial\psi_1}{\partial
\zeta}=0\,\,\,\,\,\,\mbox{for}\,\,\,\,\,\,\zeta=0.
\label{64}
  \end{equation}

Using Eqs. (\ref{43}) and integrating relations (\ref{64}) yields
\begin{equation}
\psi_1 = I^{(1)}[I^{(0)}+I^{(1)}\Gamma]\frac{1- \zeta^2}{2}
+\frac{\Omega(r)+ \phi^{(1)}}{8r^{5/2}} \big{\{}   (1-
\zeta^2)^{3/2}+\frac{2}{5} (1- \zeta^2)^{5/2} -3\big{[}\zeta
asin\zeta-\frac{\pi}{2}  +(1- \zeta^2)^{1/2} \big{]}\big{\}},
\label{65}
  \end{equation}

In that  approximation, the parameters  $I^{(0)}$,  $I^{(1)}$,
$\phi^{(1)}$ and $B$ and/or  $M$  are needed for the complete
description of the Hall equilibrium state (the parameters
$\phi^{(0)}$ and $\beta_H$ are dropped from that problem).
Solution (\ref{65})
is invariant with respect to the
simultaneous change of $I^{(0)}$, $I^{(1)}$ by $-I^{(0)}$,
$-I^{(1)}$ so that further analysis may be restricted by positive
values of  $I^{(0)}$.

It is finally remarked that in the case under consideration the
density distribution (\ref{63}) is the same as in the pure
hydrodynamic case, in particular, it vanishes at  $r\to\infty$ and
the disk is unbounded in such approximation. It may be shown that
infinite radial size disk in this approximation follows from
non-uniformity at large radius of the asymptotic expansions
(\ref{62}) in  $\delta$. Thus, according to Eq. (\ref{65})
$\psi(r,\zeta)-\delta\Gamma(r)=\delta\psi_1(r,\zeta)\sim\delta$ at
$r\to\infty$. Substituting that estimate into Eq. (\ref{41}) for
the density yields that the terms  $\sim\delta$ neglected in the
approximation (\ref{63}) are of the same order as the last term in
Eq. (\ref{41}) $\sim 1/r^3$ at sufficiently large radius  $r>>1$.
 This yields that asymptotic expansions (\ref{62})  fails for
  $r\sim \delta^{-k}$ ($k>0$) with $k$
  depending on the form of $\Gamma(r)$ and the proportionality
  coefficient depends on the Hall plasma beta parameter.
  For, instance if  $B=0$ and $\Gamma(r)=M/r$, then $r\sim\delta^{-2/3}>>1$.
 Consequently,  the disk radius   $R_d$ may be estimated
as ${R_d}\gtrsim\delta^{-k}$, ($k>0$).
A more exact determination of the disk radius (finite or infinite) requires an additional effort.
 Nevertheless, the above non-uniform solution
 may be employed with no restriction for smaller values of radial coordinate
 (e.g. $r\sim\delta^{0}$) in the local analysis of the disk stability (see e.g. \cite{Shtemler et al. 2007}).
%

\section{ NUMERICAL EXAMPLES}
The density contour lines  and the  poloidal magnetic-field lines
are calculated for linear approximations of the trial functions
$I(\psi)=I^{(0)}+I^{(1)}\psi$ and  $\phi(\psi)=\phi^{(1)}\psi$ as
well as for a model magnetic flux at the horizontal edge of the
disk $\Gamma(r)=Br^2+M/r$. The parameters $I^{(0)}$,  $I^{(1)}$,
$\phi^{(1)}$,  $B$ and/or  $M$ are chosen in such a way to cover
possible combinations of their signs  taking into account the
symmetry properties. All calculations  for equilibrium solutions
considered below may be restricted to positive values of
$I^{(0)}>0$
 with no loss of generality.
Furthermore, the parameters are selected by assuming that
for physically accepted solution density should either equal to zero at a finite radius of the disk
(first family of equilibria) or vanishing asymptotically with growing value of radius (second family of equilibria).
In all calculations  a typical value of the Hall plasma beta  $\beta_H=5$ has been fixed.
On several Figures below $\theta$ denotes a small value $\theta=0.005$.

\subsection{Numerical examples for the first family of equilibrium solutions (Figs. 1-4)}

For the first family of equilibrium solutions the disk radius  $R_d$ is exhibited as a cut off value
at which density drops to zero.
The numerical solutions are restricted to values of the parameters of the trial and edge boundary functions,
which yield  the dimensionless disk radius  $R_d$ of  order of unity.
Note that the values of the disk radius in Fig. 3b and Figs. 4b, 4d
correspond to zero net poloidal flux, and they are in the fair agreement with Eq. (\ref{61}).

The physically acceptable equilibrium solutions in Fig. 1 ( $I^{(0)}>0$) are found only
either for negative primordial magnetic field  $B<0$,  $I^{(1)}>0$, and  $\phi^{(1)}>0$,
or otherwise for  $B>0$,  $I^{(1)}<0$,  and  $\phi^{(1)}<0$.
For the dipole-like magnetic field along with qualitatively similar magnetic-field
configurations  new types appears in Fig. 2 with  quite different topology from that in Fig. 1.
The equilibrium solutions in Fig. 2 ($I^{(0)}>0$) are found only either for positive moment of magnetic dipole,
$M>0$ for  $I^{(1)}>0$ and  $\phi^{(1)}<0$  or otherwise for  $M<0$  at  $I^{(1)}<0$    and   $\phi^{(1)}>0$.
For the  case of mixed magnetic flux at the horizontal edges along with qualitatively similar
magnetic-field configurations  new types appears now in Figs. 3 and 4 with  quite different topology
from that in Figs. 1 and  2 for primordial and dipole-like at the horizontal edges.

 It is easy to distinguish two kinds of density contour lines topology:

(i) monotonically decreasing density starting from the disk center to  the disk radius  $r=R_d$
(Figs. 1-4 except of Fig. 3d ).

 (ii) A density core in the disk center that is separated from an outer density
 ring by an internal plasma-free cavity inside the disk (Fig. 3d).
 The appearance of a density ring in Fig. 3d is associated with three roots of the cubic Eq. (\ref{57}),
  which determine two lateral boundaries of the external ring and of
  the internal cavity inside which the density is identically zero (the inner
  and external radii in Fig. 3d of the disk ring are  $R_d^{(in)}$ and  $R_d$,
  and the ratio of the outer-to-inner radius is of the order of  $R_d/R_d^{(in)}\sim3$.

In Figs. 1-4 there are  two kinds  (i) and (ii) of magnetic field
structures both of which are characterized by condition
(\ref{60}), while the other two kinds (iii) and (iv) of the
magnetic field structures are obtained by employing Eq.
(\ref{59}):

 (i) Magnetic structures for zero dipole magnetic flux at the horizontal edge.
 The magnetic structures are formed by a magnetic island within the disk that
  is centered on an O-point located at the origin (Figs. 1a, 1c).
   When a magnetic line intersects the horizontal disk edges it is
   changed by the near-vertical magnetic lines that are oriented from bottom to top disk edges.

(ii) Magnetic structures for dipole-like and mixed edge magnetic fluxes.
In that case the magnetic structures are formed by near-vertical magnetic
lines oriented from bottom to top disk edges (Figs. 2a, 2c, Figs. 3a, 3c and 4a, 4c).

(iii) Magnetic structures for zero dipole and mixed edge magnetic fluxes.
In that case magnetic structures are generated which have an X-point located
either at the origin of the coordinates (Figs. 1e, 1g) or even inside the
Keplerian portion of the disk (Figs. 3e, 3g).

(iv) Magnetic structures for pure dipole-like edge magnetic flux.
In that case magnetic structures are generated which have two multivalued-points
 located in two symmetrical points with respect to the midplane on the disk axis
 ($r=0$ - not included in the Keplerian portion of the disk
 $R_d>>\epsilon$), where all magnetic-field lines  $\psi=const$ converge (Fig. 2e).

 Finally note  that the radii of the small disks are independent of the Hall parameter $\pi_H$,
 moreover they are slightly dependent  on the Hall plasma beta $\beta_H=\beta \pi_H$
 for a wide range of the parameters of the trial $I(\psi)$,  $\phi(\psi)$ and edge
 $\Gamma(r)$ functions   for which  $R_d$ is well approximated by the solution of the approximate
 Eq. (\ref{56}). Thus the values of  $R_d$ are found to be in a fair agreement with those given by
 Eq. (\ref{56}) for most of Figs. 1-4 (except of Figs. 1f and 4d where  $R_d$ is so small that the exact
 Eq. (\ref{55}) must be solved).
It is interesting  that the approximate conditions (\ref{57}) are
not satisfied for such structures as X-points as well as other
topologically unstable configurations depicted in Figs. 1h, 2f, 3f
and 3h.
%

\begin{figure*}
\includegraphics[width=120mm]{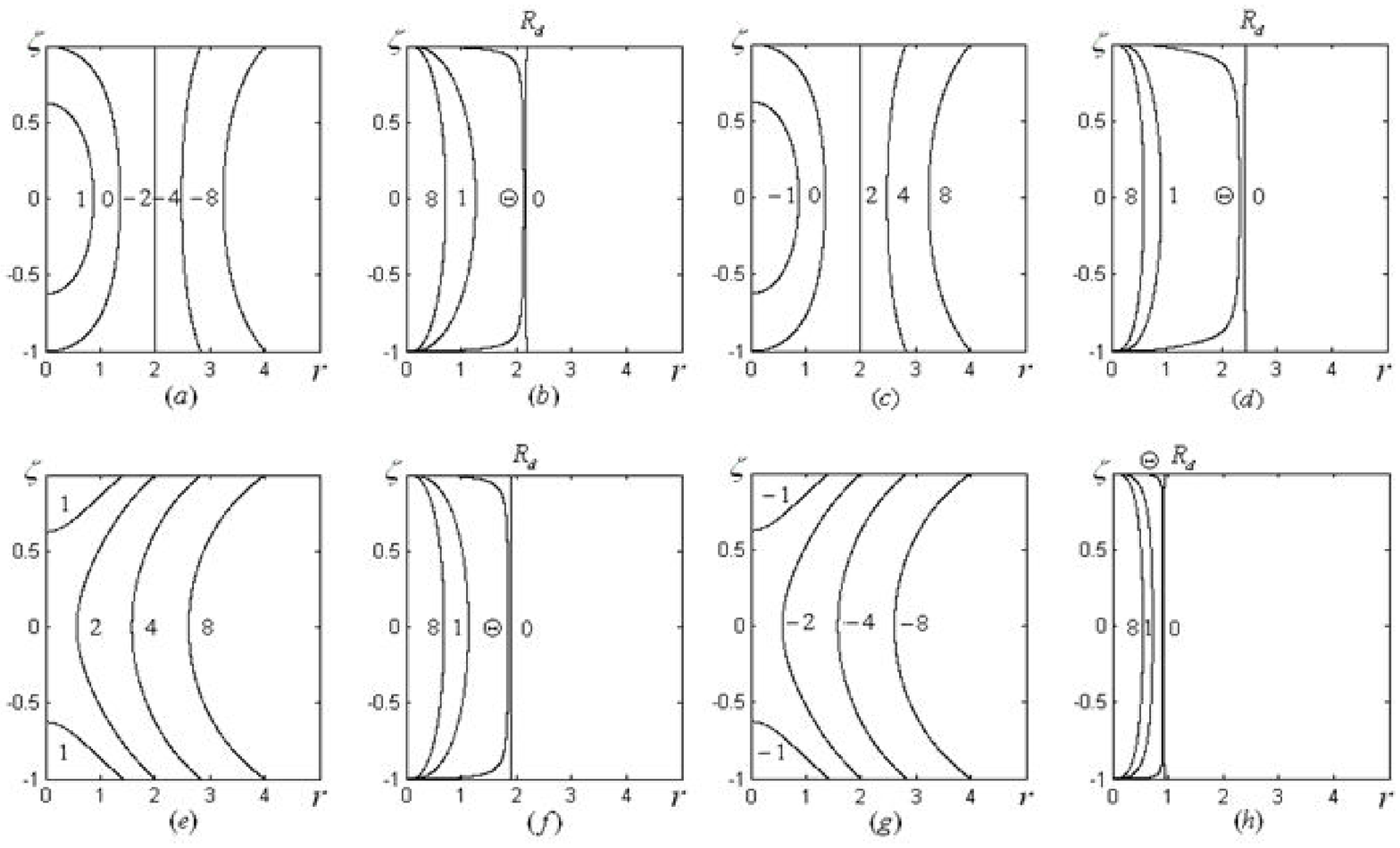}
  \caption{Magnetic field lines and density contours for pure primordial magnetic flux at the horizontal edges,
$M=0$:
  (a-b)  $B=-1$,  $I^{(0)}=2$,  $I^{(1)}=1$,  $\phi^{(1)}=0.5$, $R_{d}\approx 2.2$;
(c-d)  $B=1$,  $I^{(0)}=2$,  $I^{(1)}=-1$,  $\phi^{(1)}=-0.5$,  $R_{d}\approx2.3$;
(e-f)  $B=1$,  $I^{(0)}=2$,  $I^{(1)}=1$,  $\phi^{(1)}=-0.5$,   $R_{d}\approx1.9$;
(g-h)  $B=-1$,  $I^{(0)}=2$,  $I^{(1)}=-1$,  $\phi^{(1)}=0.5$,   $R_{d}\approx1$.
  }
\label{Fig. 1}
\end{figure*}


\begin{figure*}

\includegraphics[width=120mm]{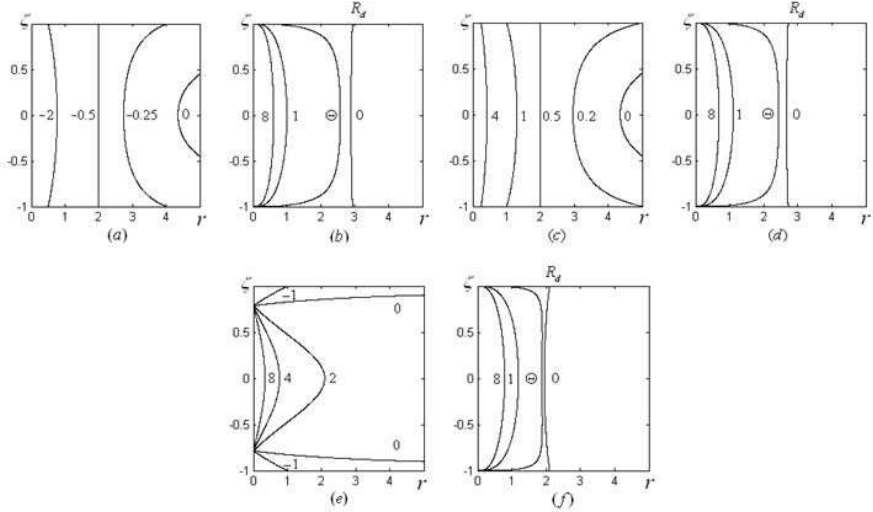}
  \caption{
  Magnetic field lines and density contours for pure dipole magnetic flux at the horizontal edges,
$B=0$:
(a-b)  $M=1$,  $I^{(0)}=0.5$,  $I^{(1)}=1$,   $\phi^{(1)}=-1$,    $R_{d}\approx2.9$;
(c-d)  $M=-1$,  $I^{(0)}=0.5$,  $I^{(1)}=-1$,  $\phi^{(1)}=1$, $R_{d}\approx2.7$;
(e-f)  $M=1$,  $I^{(0)}=0.5$,  $I^{(1)}=-2$,  $\phi^{(1)}= -0.5$,  $R_{d}\approx2$.
  }
\label{Fig. 2}
\end{figure*}


\begin{figure*}
\includegraphics[width=120mm]{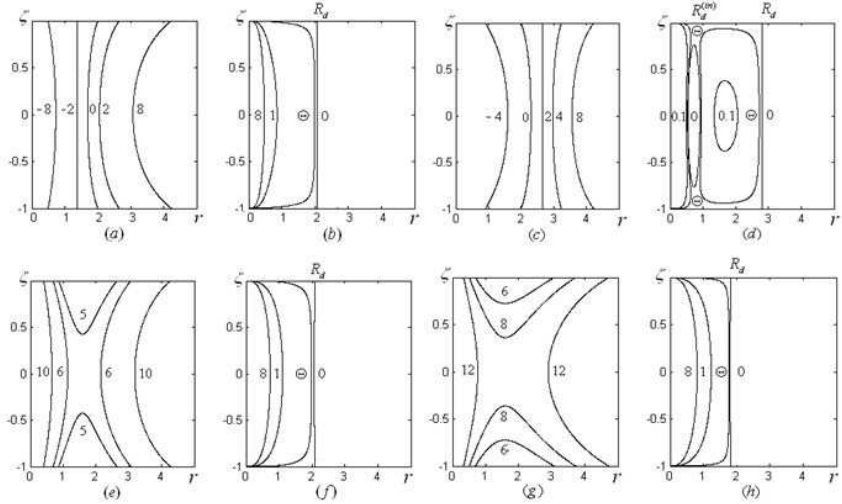}
  \caption{Magnetic field lines and density contours for mixed magnetic flux at the horizontal edges:
(a-b)  $M=4$,  $B=1$,  $I^{(0)}=2$,  $I^{(1)}=1$,  $\phi^{(1)}=-0.5$,  $R_d\approx 2$;
(c-d)  $M=4$,  $B=1$,  $I^{(0)}=2$,  $I^{(1)}=-1$,  $\phi^{(1)}=-0.5$,   $R_d\approx 2$;
(e-f)  $M=-4$,  $B=1$,  $I^{(0)}=2$,  $I^{(1)}=-1$,  $\phi^{(1)}=-0.5$,    $R_d\approx 2$;
(g-h)  $M=-4$,   $B=1$,   $I^{(0)}=2$,    $I^{(1)}=1$,  $\phi^{(1)}=-0.5$,     $R_d\approx 1.9$.
}
\label{Fig. 3}
\end{figure*}

\begin{figure*}
\includegraphics[width=120mm]{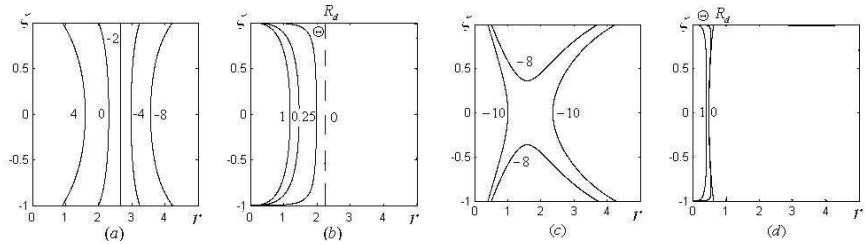}
  \caption{ Magnetic field lines and density contours for mixed magnetic flux at the horizontal edges:
(a-b)  $B=-1$,  $M=-4$,   $I^{(0)}=2$,  $I^{(1)}=1$,  $\phi^{(1)}=-0.5$, $R_{d}\approx2.2$;
(c-d)  $B=-1$,  $M=4$,   $I^{(0)}=2$,  $I^{(1)}=-1$,  $\phi^{(1)}=-0.5$, $R_{d}\approx0.5$.
}.
\label{Fig. 4}
\end{figure*}

\subsection{Numerical examples for the second family of equilibrium solutions (Figs. 5-8)}
For the second family of equilibrium solutions  the density is
given by a pure hydrodynamic Keplerian distribution (see Fig. 5),
and it is the same for all admissible magnetic field
configurations in Figs. 6-8
(with parameters $M$, $B$, $I^{(0)}$ scaled by $\delta$, and
restricted to $\phi^{(1)}>0$).
%
%
%
 %
To illustrate the qualitative behavior of the magnetic field, the
field lines of the scaled magnetic flux
$\psi_1(r,\zeta)=[\psi(r,\zeta)-\delta\Gamma(r)]/\delta$ are
depicted in Figs. 6, 7 and 8 for different kinds of the edge
magnetic flux:
with zero dipole-moment, pure dipole-like magnetic field and mixed magnetic fields. 

The second family of equilibria generates  structures that are quite similar
to those in the first family, e.g. X-points in Figs. 6c, 6d, 8f, 8g, 8h,
 and several configurations with a single O-point structures in Figs. 6a, 6b, 7b, 8c, 8d.
 In addition, new kinds of the structures arise and may be seen in Figs. 7a,7c, 7d and Fig. 8a, 8b, 8e.
 Thus, Figure 7c demonstrates a magnetic island with radius that increases infinitely with decreasing magnetic flux value.
 Then double O-point structures appears with magnetic islands additional to those centered on the O-point
 at the coordinate origin. Figures 7a, 7d and 8e demonstrate presence of two magnetic islands ("plasmoids"),
 first with central O-points located at the origin of the coordinate,
 and second with both finite radius of magnetic islands and coordinate
 of O-point inside the disk
 in Fig. 8e, or with both the radius of magnetic island and coordinate
 of O-point which tend to infinity
  in Figs.  7a and 7d.
 Figures 8a and 8b with a single O-points located at the origin separated
 from the right by a separatrix  $\psi_1=0$ demonstrate the magnetic lines with strong
 variations of their gradients from the right of the separatrix. Figure 8e demonstrates the presence of O-points,
 closed magnetic lines confined both from the left and right by the separatrix  $\psi_1=0$, while X-points
 are in  Figs. 8f-8h.
Finally note that some of the equilibrium magnetic lines obtained above are quite similar to the instantaneous
magnetic lines obtained experimentally in a tokomak (e.g. Fig. 6 in \cite{Iisuka et al. 1986}).

\begin{figure*}
\includegraphics[width=35mm]{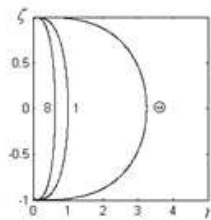}
  \caption{Density contours  for the second family of solutions.}
\label{Fig. 5}
\end{figure*}

\begin{figure*}
\includegraphics[width=120mm]{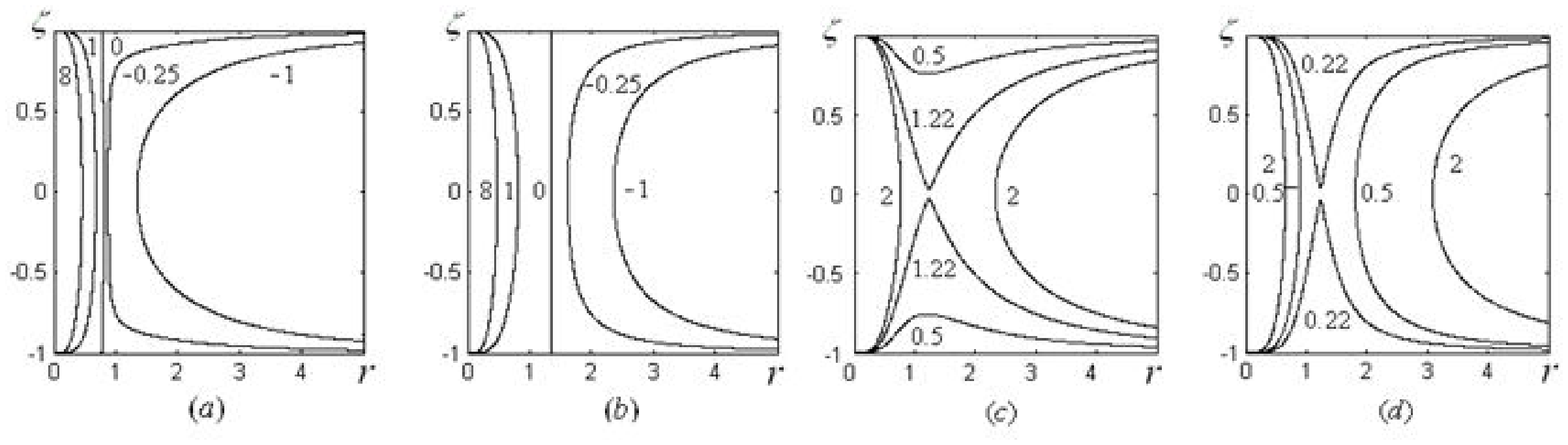}
  \caption{ Perturbed magnetic field lines for pure primordial  magnetic flux at the horizontal edges, $M=0$
, $\phi^{(1)}=1$:
(a)  $B=-1$,  $I^{(0)}=1$,  $I^{(1)}=-1$; (b)
$B=-1$, $I^{(0)}=1$,   $I^{(1)}=1$; (c)  $B=1$,  $I^{(0)}=1$,
$I^{(1)}=1$; (d)  $B=1$,  $I^{(0)}=1$,  $I^{(1)}=-1$.
  }
\label{Fig. 6}
\end{figure*}


\begin{figure*}
\includegraphics[width=120mm]{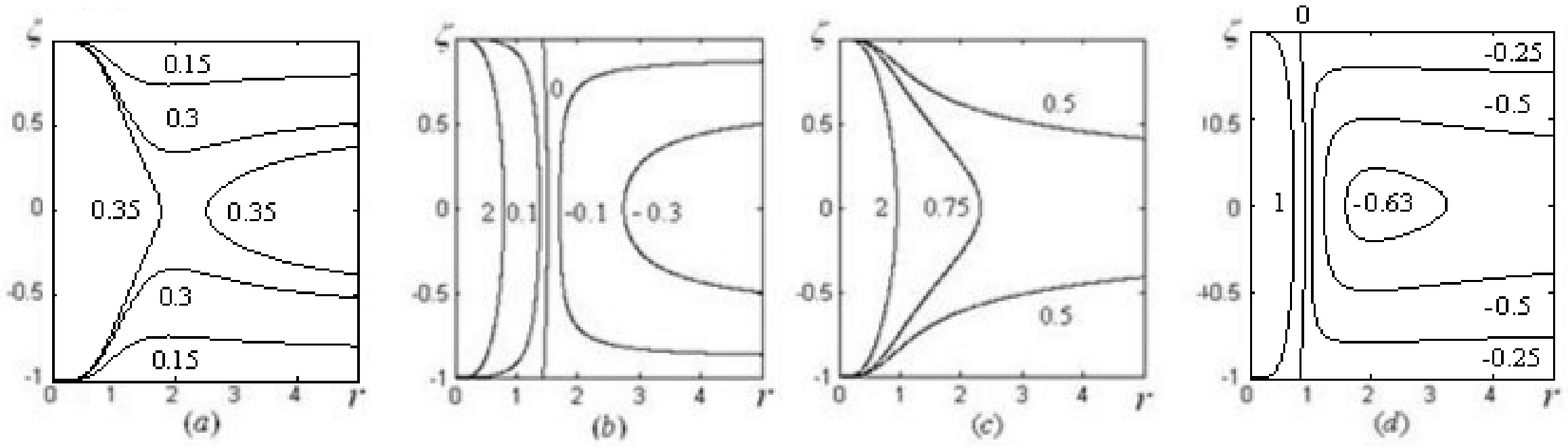}
  \caption{Perturbed magnetic field lines for pure dipole  magnetic flux at the horizontal edges, $B=0$, $\phi^{(1)}=1$:
(a)  $M= 1$,  $I^{(0)}=1$,  $I^{(1)}=1$; (b)  $M= -1$,
$I^{(0)}=1$,  $I^{(1)}=-1$; (c)  $M= -1$,  $I^{(0)}=1$,
$I^{(1)}=1$; (d)  $M= 1$,  $I^{(0)}=1$,  $I^{(1)}=-1$.
  }
\label{Fig. 7}
\end{figure*}


 \begin{figure*}
\includegraphics[width=120mm]{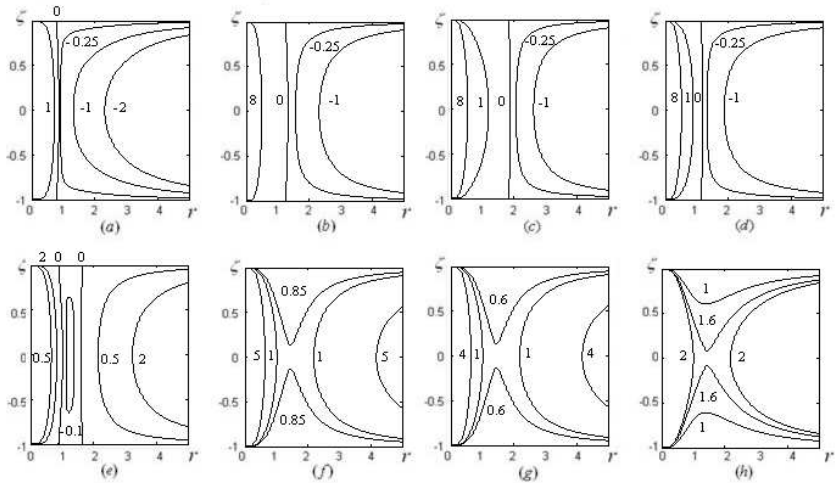}
  \caption{Perturbed magnetic field lines for mixed magnetic flux at the horizontal edges, $\phi^{(1)}=1$:
(a)  $B=-1$,  $M=1$,   $I^{(0)}=1$,   $I^{(1)}=-1$; (b)  $B=-1$,
$M=1$,   $I^{(0)}=1$,   $I^{(1)}=1$; (c)  $B=-1$,  $M=-1$,
$I^{(0)}=1$,   $I^{(1)}=1$; (d)  $B=-1$,  $M=-1$,  $I^{(0)}=1$,
$I^{(1)}=-1$; (e)  $B=1$,  $M=1$,    $I^{(0)}=1$,   $I^{(1)}=-1$;
(f)  $B=1$,  $M=1$,    $I^{(0)}=1$,   $I^{(1)}=1$; (g)  $B=1$,
$M=-1$,   $I^{(0)}=1$,   $I^{(1)}=-1$; (h)  $B=1$,  $M=-1$,
$I^{(0)}=1$,   $I^{(1)}=1$.
  }
\label{Fig. 8}
\end{figure*}

\section{SUMMARY AND DISCUSSION}
3D axially-symmetric steady-state equilibria of weakly ionized polytropic Hall-MHD plasmas are investigated for the Keplerian portion of thin disks. Asymptotic expansions in small aspect ratio  $\epsilon$ provide an efficient way to construct an equilibrium model for differentially rotating disks. There are three principle Hall equilibrium states which can not be reduced from one to the other, classified according to the following orientations of the magnetic field lines: (i) pure toroidal, (ii) pure poloidal, and (iii) mixed toroidal and poloidal equilibria. The three classes of equilibria differ from each other by the different ordering with respect to $\epsilon$  of the various components of the magnetic field. In the present study the most interesting case is investigated that involves both toroidal and poloidal components of the magnetic field that are of the same zero order in $\epsilon$. The disk structure is described by the GS equation for the poloidal flux   that involves two arbitrary functions  $I(\psi)$ and  $\phi(\psi)$  for the toroidal electric current and electric potential, respectively.
The arbitrariness of functions $I(\psi)$ and $\phi(\psi)$  in steady-state problems reflects the uncertainty of the initial data in the general initial value problem.
The flux function   is symmetric about the midplane and satisfies certain boundary conditions at the horizontal edges of the disk (radial variations of the disk height are neglected).
The boundary conditions for the magnetic flux $\psi=\Gamma(r)$
at the horizontal disk edges express the joint effect of an primordial magnetic field, and a dipole-like magnetic field that reflects the influence of the central body on the Keplerian disk.
Solutions for different configurations of the magnetic field and density are obtained explicitly for trial linear approximations of the current  $I(\psi)=I^{(0)}+ I^{(1)}\psi$  flowing through a circular area in a plane   and of the electric potential   $\phi(\psi)=\phi^{(1)}\psi$   (where   $I^{(0)}$ is the total current concentrated along the disk axis), as well as several forms of the boundary magnetic flux $\Gamma(r)$.

A particular noteworthy new feature of the present model is the finite radius of the rotating disks which is inherent for the Hall equilibrium. By using a small parameter  $\delta$  proportional to the inverse Hall parameter with a small  coefficient of proportionality, it is established that the Hall equilibria disks fall into two types which are characterized by quite different orders in $\delta$: Keplerian disks   with (i) small ($R_d\sim\delta^0$) and (ii) large (${R_d}\gtrsim\delta^{-k}$, $k>0$) radius of the disk.
For disks of the first family a finite radius of the disk  appears as a cut-off value at which the plasma density vanishes.
Disks of the second family  have  large  (finite or infinite)
radii  due to non-uniformity of the asymptotic expansions in
$\delta$.

%
The method developed here allows to investigate analytically the equilibrium states with a large number of possible combinations of boundary conditions for the magnetic flux. Thus, it is demonstrated that all possible mutual orientations of the rotating axis, the primordial magnetic field, and the magnetic moment of the central body, as well as the relative strength of the two latter give rise to a great richness of possible topologies of the magnetic field lines. Such configurations includes geometries with O-points, with X-points, with material gaps and rings, with detached plasmoids, with magnetic islands, and with zero net induced magnetic flux.
Note that magnetic islands centered on O-points are typical for steady-state equilibria in Keplerian disk. For instance they have been simulated in thin disk approximation within MHD model for the pure poloidal magnetic field (\cite{Lovelace et al. 1986}). On the other hand,  equilibria that contain X-points are more subtle and generally have a strong tendency to undergo a significant change in their topology due to possible reconnection processes (\cite{Priest and Forbes 2000}). The latter however is commonly speculated to be responsible for widely observed astrophysical phenomena like jets and winds.
%
%
Describing such processes requires an extensive stability analysis. Since the primordial poloidal flux on the disk is likely too small to be responsible for the observed total magnetic flux in disk systems (Colgate and Li 2001), the possible flux enhancing due to instability of axisymmetric Hall equilibria becomes an important issue. Additionally, it is natural then to incorporate multipoles of higher odd orders (Lepeltier and Aly 1999) instead of the primordial poloidal flux, in addition to the dipole adopted here, into to the boundary flux sources at the origin that describe the central object effect on the Keplerian disk. The stability study of the developed equilibrium disks may provide further selection of admissible magnetic configurations. For instance, development of X-point configurations may lead to an instability of such equilibria, since X-point equilibrium solutions are rather structurally unstable and a small perturbation of the equilibrium likely leads to an essential change in its form, such as magnetic lines reconnection and transition to the neighborhood stable (smooth) configuration. The present Hall equilibrium solution for thin Keplerian disks may be incorporated to the leading order in small aspect ratio into a more sophisticated problem accounting for accretion and jet inflow (similar to Ogilvie 1997).

\section*{Acknowledgments}
The authors thank Edward Liverts for useful and insightful discussions.

\bsp
\label{lastpage}
\end{document}